\documentclass[12pt]{iopart}
\usepackage{amsbsy,amssymb,xspace,enumerate,color,mathptmx}
\makeatletter
\let\fref\@undefined
\let\Fref\@undefined
\makeatother
\usepackage[plain]{fancyref}
\usepackage[utf8]{inputenc}
\usepackage{graphicx}
\usepackage{wrapfig}
\newcommand{\ie}{\textit{i.\,e.},\ }

\newcommand{\cf}{\textit{cf.}\ }
\newcommand{\HBM}{hot Brownian motion\xspace}

\newcommand{\THBM}{T_\mathrm{HBM}^\mathrm{r}}
\newcommand{\zetaHBM}{\zeta_\mathrm{HBM}^\mathrm{r}}
\newcommand{\DHBM}{D_\mathrm{HBM}^\mathrm{r}}
\renewcommand{\vec}[1]{\mathbf{#1}}
\newcommand{\w}{w} 
\newcommand{\tensor}[1]{\mathsf{#1}}
\newcommand{\eqs}[1]{\fancyrefeqlabelprefix s.\fancyrefdefaultspacing(#1)}

\newcommand{\inst}[1]{$^{#1}$}
\newcommand{\text}[1]{$#1$}
\newcommand{\eqref}[1]{(\ref{#1})}
\definecolor{mypurple}{RGB}{153,61,113}
\definecolor{myblue}{RGB}{63,61,153}
\definecolor{myokker}{RGB}{153,140,61}
\definecolor{mygreen}{RGB}{61,153,86}
\definecolor{mymarine}{RGB}{61,90,153}
\definecolor{mycyan}{RGB}{0,255,255}

\begin{document}
\title{Rotational hot Brownian motion}
\author{D Rings\inst{1}, D Chakraborty\inst{1,2} and
  K Kroy\inst{1}}

\address{\inst{1} Institute for Theoretical Physics -- University of
  Leipzig}
\ead{klaus.kroy@uni-leipzig.de}
\address{\inst{2} \textsl{present address:} Theory of Inhomogeneous
  Condensed Matter -- Max-Planck-Institute for
  Intelligent Systems, Stuttgart}

\begin{abstract}
  We establish an effective Markov theory for the rotational Brownian
  motion of hot nanobeads and nanorods. Compact analytical expressions
  for the effective temperature and friction are derived from the
  fluctuating hydrodynamic equations of motion. They are verified by
  comparison with recent measurements and with GPU powered parallel
  molecular dynamics simulations over a wide temperature range. This
  provides unique insights into the physics of hot Brownian motion and
  an excellent starting point for further experimental tests and
  applications involving laser-heated nanobeads, nanorods and Janus
  particles.
\end{abstract}
\maketitle

\section{Introduction}
\label{sec:introduction}

The popular Markovian theory of Brownian motion, as developed by
Einstein, Langevin, and Smoluchowski a century ago, has been the
starting point and inspiration for innumerable applications
\cite{frey-kroy:2005,Haenggi:2005}. However, the usual convenient
formulation in terms of the centre-of-mass coordinates of particles
only pertains to the special case of an isolated spherical
particle. In the general case of interacting and/or anisotropic
particles, both translational and rotational degrees of freedom
couple, calling for a more elaborate mathematical description. This is
most obvious for rod-shaped particles that have different mobilities
for the movement parallel and perpendicular to their long axis
\cite{Doi:1988}, but in fact also holds for interacting spherical
particles \cite{Martin:2006}. Due to the associated technical
complications, the present theoretical understanding is still
relatively incomplete \cite{Hagen:2011}, in particular with regard to
micro-swimmers and other active or self-propelled colloidal particles
\cite{Golestanian:2009,Valadares:2010,Popescu:2010}, for which the
proper hydrodynamic description is even more subtle than for passive
particles in external fields
\cite{julicher-prost:2009,Brady:2011}. The directed motion for such
self-propelled particles from sperms \cite{Friedrich:2009} to Janus
particles running on chemical fuel \cite{Howse:2007} is usually
limited by (equilibrium or nonequilibrium) rotational Brownian
motion. Besides, rotational Brownian motion is undoubtedly of interest
for its own sake. It is accessible to spectroscopy \cite{Loman:2010}
and has been the basis for the development of new microrheological
techniques \cite{Cheng:2003} and nanoscopic heat engines
\cite{Filliger:2007}.

In this paper, we are concerned with a specific type of rotational
Brownian motion that occurs whenever the colloidal particles have an
elevated temperature with respect to their solvent. In this case, we
speak of \emph{rotational hot Brownian motion}, in analogy to the
better understood translational case \cite{Rings:2010}. Both intended
\cite{Jiang:2010,Lasne:2006,vanDijk:2006,octeau-etal:2009,Gaiduk:2010}
and unintentional \cite{Peterman:2003,Ruijgrok:2011} realizations of
(rotational) hot Brownian motion are nowadays widespread in
biophysical and nanotechnological applications, which often employ
nanoparticles exposed to laser light as tracers, anchors, and
localized heat sources. Deliberate heating of nanoparticles is, for
instance, common in photothermal therapy \cite{Chen:2010,Huang:2010},
but it also helps to enhance the optical contrast for detection
\cite{Berciaud:2004} or in photothermal correlation spectroscopy
\cite{radunz-etal:2009}. Laser-heating is also a convenient way of
supplying the energy for the self-thermophoretic propulsion of
anisotropic particles \cite{Jiang:2010}. A quantitative theory for the
optical scattering from dissolved hot nanoparticles has only recently
become available \cite{Selmke:2012}, paving the way for a broad range
of future applications.

In the following, we show that the heating affects the rotational and
translational degrees of freedom differently, which is due to the
intrinsic nonequilibrium nature of the phenomenon. To this end, we
perform analytical calculations based on nonequilibrium fluctuating
hydrodynamics, which we compare to large-scale molecular dynamics (MD)
simulations of an atomistic model of a nanoparticle dissolved in a
Lennard-Jones fluid. Although, in practice, even for an isolated
single colloid rotational and translational diffusion always occur
simultaneously, we may focus on the one or the other separately, in
theory. The requirement for this considerable simplification is that
the coupling only results in a superposition of the respective
displacements in space and orientation. To be more specific, we assume
that the conditions governing the rotational Brownian motion, \ie the
spatially heterogeneous solvent viscosity and temperature around the
particle, do not depend on the translational Brownian motion.
\begin{wrapfigure}{r}{0.45\textwidth}
  \setlength\mathindent{0cm}
  \centering
  \includegraphics[width=0.45\textwidth]{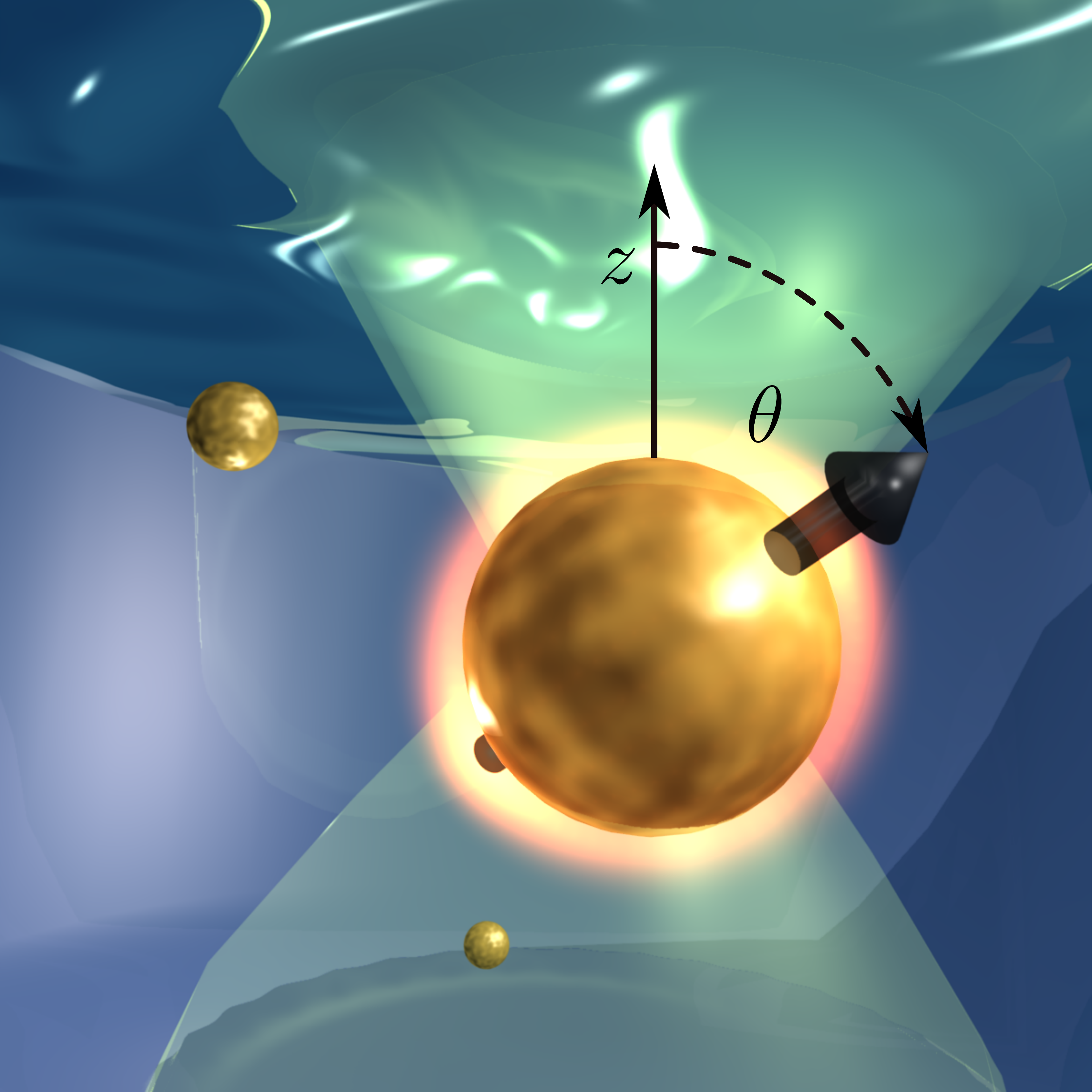}
  \caption{Artist's conception of a hot Brownian particle, 
  illustrating some notation.}
  \label{fig:artistic_sketch}
\end{wrapfigure}
This assumption relies on the common Brownian scale
separation. Typical nanoparticle diffusivities are on the order of
$10^{-11}\ldots 10^{-10}\,\mathrm{m}^2\,\mathrm{s}^{-1}$ while heat
and vorticity diffuse at
$10^{-7}\,\mathrm{m}^2\,\mathrm{s}^{-1}$. Thus, the hot Brownian
motion of a single spherical nanobead constitutes a stationary
nonequilibrium process with fixed radial temperature and viscosity
fields, $T(r)$ and $\eta(r)$, respectively, if the instantaneous
particle position is taken as the origin of the coordinate system. On
this basis, we construct and validate a Markov model for the
rotational Langevin dynamics of a hot Brownian particle with effective
temperature and friction parameters, $\THBM$ and $\zetaHBM$. While the
success of this strategy has already been demonstrated for the
translational motion \cite{Rings:2010,Rings:2011,Chakraborty:2011},
recent experiments using heated nanorods \cite{Ruijgrok:2011} and hot
Janus particles \cite{Jiang:2010} underscore the need for a separate
quantitative analysis of the rotational dynamics.  Below, we derive
$\THBM$ and $\zetaHBM$ for the rotational Brownian motion of a hot
particle and demonstrate that they differ from their analogues for
translational motion. The mathematical structure of the rotational
dynamics is simpler and allows for analytical solutions where one has
to resort to numerical methods in the translational case.

The effective Langevin equation for the rotational dynamics of the nanoparticle 
orientation $\vec{n}$ reads \cite{Kalmykov:1996},
\begin{equation}
  \label{eq:Langevin_eq_BP}
  \zetaHBM\dot{\vec{n}}=\boldsymbol{\xi}\times\vec{n}\;.
\end{equation}
Under the presupposed nonequilibrium steady-state conditions outlined
above, $\zetaHBM$ is the effective rotational friction coefficient. In
the isothermal limit, $\zetaHBM\to 8\pi \eta R^3$ for a sphere of
radius $R$ in a solvent of viscosity $\eta$. The stochastic torque
$\boldsymbol{\xi}$ is assumed to be a Gaussian random variable
characterized by the moments
\begin{equation}
  \label{eq:fluctuating_force}
  \langle\boldsymbol{\xi}(t)\rangle=0\;,\quad\langle\xi_i(t)\xi_j(t')\rangle=
  2k_\mathrm{B}\THBM\zetaHBM\delta_{ij}\delta(t-t')\;.
\end{equation}
The form of the noise strength amounts to the assumption that a
generalized Einstein relation
\begin{equation}
  \label{eq:GER}
  k_\mathrm{B}\THBM=\DHBM \,\zetaHBM
\end{equation} 
links the effective friction and temperature, $\zetaHBM$ and $\THBM$,
to the effective rotational diffusivity $\DHBM$. That the Brownian
dynamics of a single heated particle is indeed constrained by this
quasi-equilibrium relation is corroborated by our MD simulations,
presented below.

\section{Theory}
The classical rotational Stokes problem is to find the friction
coefficient of a steadily rotating particle in a viscous fluid of
homogeneous solvent viscosity. For heated particles, the assumption of
a constant viscosity has to be relaxed. The general case of an
arbitrarily shaped hot particle, which induces an asymmetric
temperature profile in the solvent, gives rise to formidable technical
complications. With the aim of deriving analytical results, we
restrict our discussion to spherical beads (\cf
\fref{fig:artistic_sketch}). Then the temperature can be idealized as
a radial field $T(r)$ that entails a radially varying viscosity
$\eta(r)$ via some constitutive law of the solvent, which we assume to
be given. Some extensions to spheroids and slender cylinders can be
discussed along the same lines if the radius $r$ is given a slightly
different interpretation, as outlined below \cite{Avudainayagam:1972}.
We further take the solvent to be incompressible, which is a good
approximation for most common solvents, such as water, and eases the
calculation. In the low Reynolds number limit, applicable to micro-
and nanoparticles in solution, the solvent velocity $\vec{u}(\vec{r})$
follows from
\begin{equation}
  \label{eq:Stokes_eq}
  \nabla p = 2\nabla\cdotp\eta\tensor{\Gamma} = 2\tensor{\Gamma}\nabla\eta+2\eta\nabla\cdotp\tensor{\Gamma}\,,\quad\nabla\cdotp\vec{u}=0\;,
\end{equation}
with the strain rate tensor
$\tensor{\Gamma}\equiv(\nabla\vec{u}+\nabla\vec{u}^T)/2$ and the pressure $p$.  

For the purely rotational fluid motion around a steadily rotating
sphere, analytical solutions can be found, as follows. The high
symmetry of the velocity field
\begin{equation}
  \label{eq:velocity_ansatz}
  \vec{u}(r,\theta,\phi) = u_\phi\vec e_\phi=\w(r)\sin\theta\,\vec{e}_\phi
\end{equation}
around the sphere entails a highly degenerate strain rate tensor. The
tensor element $\Gamma_{\theta\phi}$ contains $u_\phi$ only in the
form $\partial_\theta(u_\phi/\sin\theta)=\partial_\theta\w(r)=0$,
leaving us with only a single relevant entry $\Gamma_{\phi r}$ with
\begin{equation}
  \label{eq:strain_tensor}
  2\Gamma_{\phi r} =
 \partial_ru_\phi- u_\phi/r =
 (\w'-\w/r)\sin\theta \;.
\end{equation}
Since the viscosity is assumed to vary only radially, its gradient is
$\nabla\eta(r)=\eta'\vec{e}_r$, and the equation of motion,
\fref{eq:Stokes_eq}, reduces to
\begin{equation}
  \label{eq:NS_simple}
    \nabla p = \left[2\eta'\Gamma_{\phi r} + \eta\nabla^2
      u_\phi-\eta u_\phi r^{-2}\sin^{-2}\theta\right]\vec e_\phi
\end{equation}
Due to the cylindrical symmetry, the pressure gradient must not
contain an azimuthal component, though, \ie $\nabla p=0$. This leads
to the ordinary differential equation for $\w(r)$,
\begin{equation}
  \label{eq:general_ode}
  \w''+ (2/r+\eta'/\eta)(\w'-\w/r)= 0\,.
\end{equation}
For the known case of a constant viscosity $\eta(r)=\eta_0$ one easily
verifies by insertion the solution $\w=c_1r+c_2/r^2$ with constants
$c_1$ and $c_2$. For an unbounded fluid $c_1=0$ so that the flow field
takes the familiar form $\vec
u=\boldsymbol{\Omega}\times\vec{r}(R/r)^3$ for a sphere of radius $R$
rotating at constant angular velocity $\boldsymbol{\Omega}$.  A
constant viscosity can be interpreted as a degenerate case (for $n=2$)
of the power law viscosity field
\begin{equation}
  \label{eq:power_law_visc}
  \eta(r)=\eta_0(r/R)^{n-2}\,.
\end{equation}
For the latter $\eta'/\eta=(n-2)/r$, hence \fref{eq:general_ode}
reduces to $\w''+n\w'/r-n\w/r^2=0$, which has the general solution
$\w(r)=c_1r+c_2r^{-n}$.  Thus the flow field in an unbounded fluid
with a power-law viscosity field with arbitrary $n>0$, reads
\begin{equation}
  \label{eq:arb_power_solution}
  \vec u=\boldsymbol{\Omega}\times\vec{r}\,(R/r)^{n+1}\,.
\end{equation}
For more complex viscosity profiles $\eta(r)$, the task of solving
\fref{eq:general_ode} can be reduced to an integration using the
Wronskian. Knowing the particular solution $\w(r)=r$, the general
solution to \fref{eq:general_ode} is found to be
\begin{equation}
  \label{eq:gensol_gen_problem}
  \w(r)=c_1r+c_2r\int^\infty_r\frac{1}{\eta(x)x^4}\,\rmd x\,.
\end{equation}
Moreover, rewriting \fref{eq:general_ode}  in terms of
$\gamma \equiv \w'-\w/r$ yields 
\begin{equation}
  \label{eq:psi_equation}
  \gamma'+([\ln\eta]'+3/r)\gamma=0\,,
\end{equation}
which is (up to a constant factor) solved by
\begin{equation}
  \label{eq:psi_solution}
  \gamma(r)\propto (R/r)^3\!/\eta(r) \;.
\end{equation}

With the solution of the Stokes problem at hand, we can explicitly
calculate the effective friction coefficient $\zetaHBM$ for a rotating
sphere. The torque exerted by a rotating sphere on the surrounding
fluid is obtained by integrating $2\eta\Gamma_{\phi
  r}\vec{r}\times\vec{e}_\phi =
\eta\gamma\sin\theta\,\vec{r}\times\vec{e}_\phi$ over the surface of
the sphere, where
$\vec{r}\times\vec{e}_\phi=-R\,\vec{e}_\theta=R\sin\theta\,\vec{e}_z-R\cos\theta\,\vec{e}_\rho$. As
the $\phi$-integral over the $\vec{e}_\rho$-component vanishes
identically, this yields
\begin{equation}
  \label{eq:torque2}
  2\pi\int_0^\pi \rmd\theta \;
  R^3\sin^3\theta\eta(R)|\gamma(R)|=\frac{8\pi}{3}
  R^3\eta(R)|\gamma(R)| \;,
\end{equation}
and division by $|\boldsymbol{\Omega}|$ yields the wanted effective
friction coefficient.  The boundary conditions
$\vec{u}(\vec{r})|_{|\vec{r}|=R}$ and
$\lim_{|\vec{r}|\to\infty}\vec{u}(\vec{r})$, and the viscosity profile
$\eta(r)$ fix the constants $c_1$ and $c_2$ in
\fref{eq:gensol_gen_problem}. For an unbounded medium,
$w(r\to\infty)=0$ implies $c_1=0$, and $w(r=R)=\Omega R$ fixes
$c_2$. Collecting results, we find
\begin{equation}
  \label{eq:zeta_final}
  (\zetaHBM)^{-1} = \frac{3}{{8\pi}}
  \int_R^\infty\!\!\frac{1}{\eta(r) r^{4}}\,\rmd r \,,
\end{equation}
which indeed attains the isothermal value $1/(8\pi \eta_0 R^3)$ for
$\eta(r)=\eta_0$.

Following the derivations in Ref.~\cite{Chakraborty:2011}, the second
important parameter of a hot Brownian particle, its effective Brownian
temperature $T_\mathrm{HBM}$, is given by
\begin{equation}
  \label{eq:T_HBM}
  T_\mathrm{HBM}=\frac{\int_V T(\vec{r})\phi(\vec{r})\,\rmd^3
    r}{\int_V \phi(\vec{r})\,\rmd^3 r }\,.
\end{equation}
(Depending on the type of motion of the particle, the result gives the
translational/rotational effective temperature
$T_\mathrm{HBM}^\mathrm{t,r}$.) It can explicitly be determined by
integration once the dissipation function
$\phi\equiv\eta\,\tensor{\Gamma}\!:\tensor{\Gamma}/2$ is known. For
the rotational flow field in \fref{eq:velocity_ansatz},
\begin{equation}
  \label{eq:dissipation_function}
  \phi=\eta\left(\w'-\w/r\right)^2\sin^2\theta\equiv\eta\gamma^2\sin^2\theta\,,
\end{equation}
from which we get
\begin{equation}
  \label{eq:T_HBM_rot_final}
  \THBM = 
 \frac{\int_R^\infty T(r)\eta^{-1}(r)r^{-4}\,\rmd r}
  {\int_R^\infty \eta^{-1}(r)r^{-4}\,\rmd r}\,.
\end{equation}
Together with \fref{eq:zeta_final}, this completes our formal
derivation of the effective friction and temperature parameters
characterizing the rotational hot Brownian motion of a sphere. They
can explicitly be evaluated and used in the effective Langevin
\fref{eq:Langevin_eq_BP} \& (\ref{eq:fluctuating_force}), provided
that the temperature dependence $\eta(T)$ of the solvent viscosity and
the temperature profile $T(r)$ around the nanoparticle are known. For
the practically important special case that the solvent is water,
which is well characterized by
\begin{equation}
  \label{eq:VF}
  \eta(T)=\eta_\infty e^{A/(T-T_\mathrm{VF})} \text{ and } T(r)=T_0 + 
  \Delta T R/r \,,
\end{equation}
the convenient approximation (accurate to within $2\%$ for $\Delta
T\lesssim T_0$)
\begin{equation}
  \label{eq:approx_THBM}
  \THBM\approx T_0+\frac{3}{4} \Delta T
\end{equation}
is obtained by neglecting the temperature dependence of the viscosity,
\ie by setting \mbox{$\eta(r) =$ \textsl{constant}}, in
\fref{eq:T_HBM_rot_final}.

\begin{figure*}
  \includegraphics[width=0.49\linewidth]{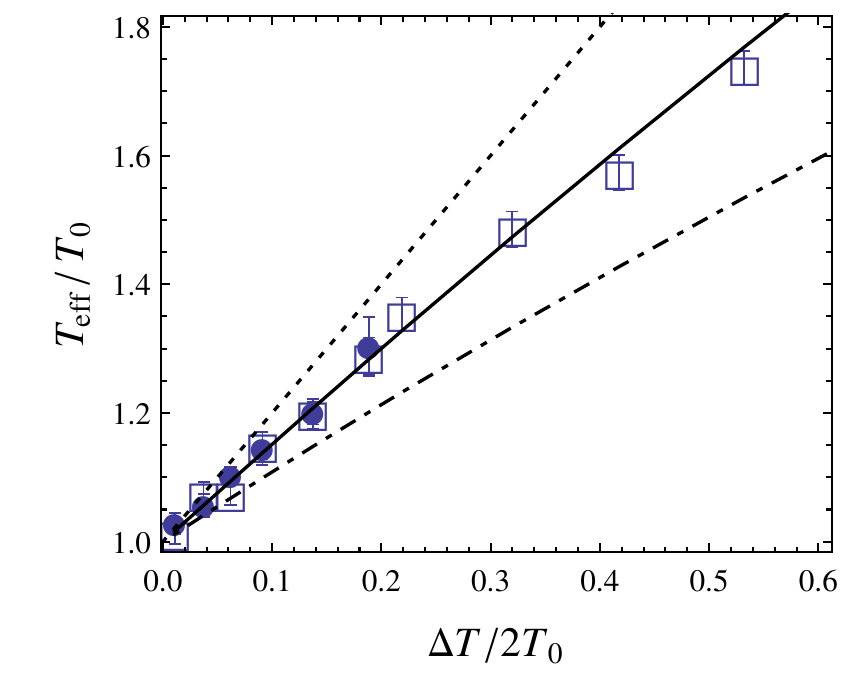}
  \hfill
  \includegraphics[width=0.49\linewidth]{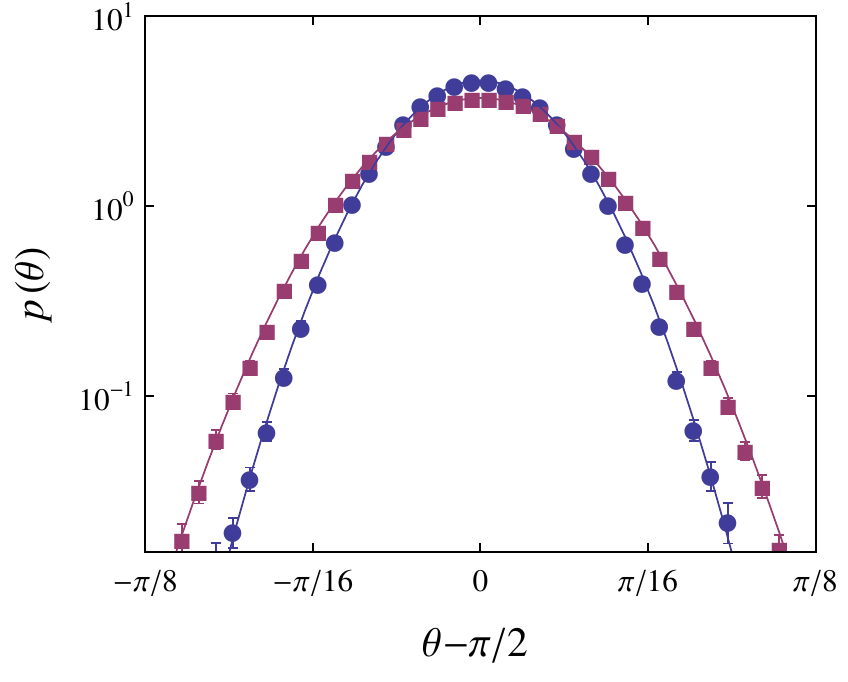}
  \caption{\textbf{Left:} Effective temperature of rotational
    \HBM. The simulation results for $\THBM$ ({\large \color{myblue}
      $\bullet$}) were deduced from the numerically measured
    $\zetaHBM$ and $\DHBM$ using the generalized Einstein relation
    \eqref{eq:GER}. An alternative estimation of $\THBM$
    ({\color{myblue}\footnotesize $\square$}) was obtained from the
    Boltzmann distribution of the inclination angle $\theta$ in a
    harmonic angular confinement (right panel). The theoretical
    prediction (solid line) was evaluated within the idealized theory
    for an incompressible fluid via \fref{eq:T_HBM_rot_final} using
    the radial viscosity profile $\eta(r)$ determined in the MD
    simulation. For comparison, the effective temperature
    $T_\mathrm{HBM}^\mathrm{t}$ for the translational degrees of
    freedom (dot-dashed line) \cite{Chakraborty:2011} and the solvent
    temperature at the particle surface (dotted line) are shown.
    \textbf{Right:} The measured distribution of the inclination angle
    $\theta$ in a harmonic angular confinement potential for
    nanoparticle temperatures $T_\mathrm{p}
    =0.8\,\epsilon/k_\mathrm{B}$ ({\color{myblue} {\large
        $\bullet$}}), $1.25\,\epsilon/k_\mathrm{B}$
    ({\color{mypurple}\footnotesize $\blacksquare$}) and the
    corresponding distribution $p(\theta) \sim e^{-\beta V(\theta)}$
    with $\beta^{-1}=k_\mathrm{B}\THBM$ depicted by the solid lines.}
  \label{fig:fig4}
\end{figure*}

For the example of a Lennard-Jones fluid with an immersed hot
nanobead, we have evaluated \fref{eq:T_HBM_rot_final} numerically,
based on the temperature and viscosity profiles obtained from our
simulations.  Care has been taken to include the density variations
and finite size effects correctly, as detailed in
Ref.~\cite{Chakraborty:2011}. In \fref{fig:fig4}, we compare the
predicted $\THBM$ to the effective temperature deduced from the
directly measured friction $\zetaHBM$ and diffusivity $\DHBM$ via the
generalized Einstein relation \eqref{eq:GER}. (A brief description of
the MD simulations can be found further below.)

\begin{figure}
  \includegraphics[width=0.49\linewidth]{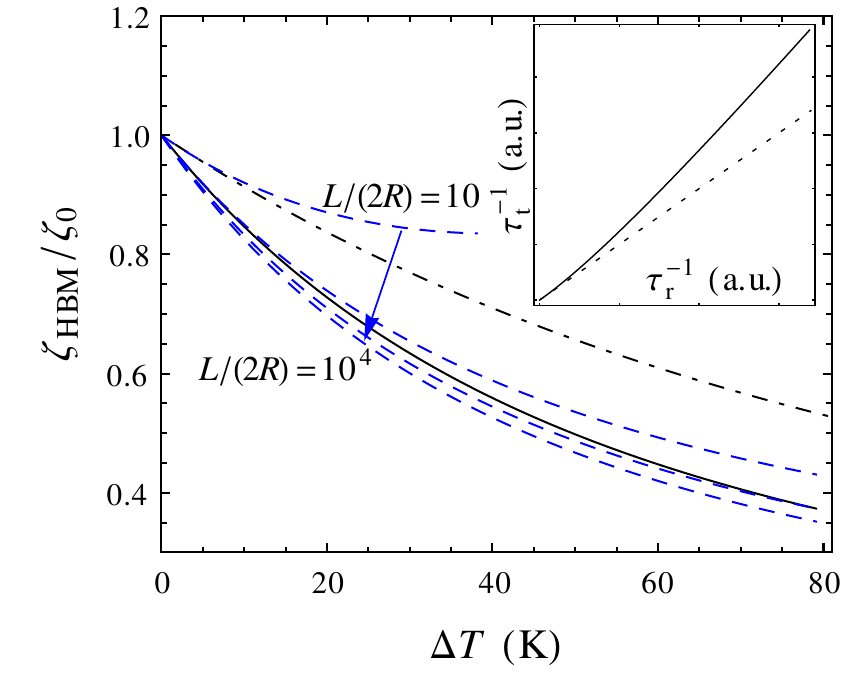}
  \hfill
  \includegraphics[width=0.49\linewidth]{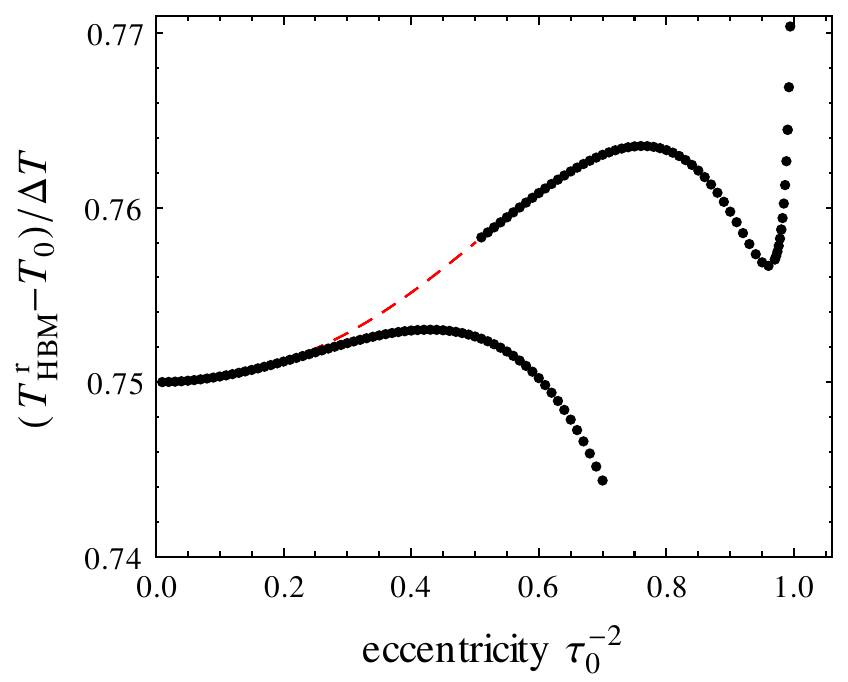}
  \caption{\textbf{Left:} Effective friction coefficient
    $\zeta_\mathrm{HBM}$ of a hot nanoparticle in water (normalized to
    the isothermal limiting value) as a function of the temperature
    increment $\Delta T$ above the ambient temperature
    $T_0=298\,$K. The analytical predictions for a spherical bead of
    radius $R$ from \fref{eq:zeta_final} and Ref.~\cite{Rings:2011}
    are represented by solid and dot-dashed lines for rotational and
    translational motion, respectively. Note that they differ by a
    heating-dependent kinematic factor, an effect that we neglect in
    the estimate \fref{eq:zeta_perp}, for slender rods. Dashed lines
    represent the effective longitudinal friction coefficient
    $\zeta^\parallel_\mathrm{HBM}$ of a hot slender rod according to
    \fref{eq:zeta_rod_approx}, for the aspect ratios $L/(2R)=10^1$,
    $10^2$, $10^3$, $10^4$.  The curves closely follow the prediction
    for the rotating sphere, except for small $L/(2R)\lesssim 10^2$
    near the isotropic limit, which is not correctly recovered by the
    slender-rod approximation.
    %
    %
    \emph{Inset:} The inverse translational versus the inverse
    rotational relaxation time of a spherical particle in an optical
    trap of varying strength; to be compared with Fig.~3 d) of
    Ref.~\cite{Ruijgrok:2011}. The dotted line indicates the naive
    estimate obtained by identifying the effective viscosities for
    translational and rotational motion. \textbf{Right:} Variation of
    the first order coefficient of $\THBM(\Delta T)$ with the
    eccentricity $\tau_0^{-2}$ of the spheroidal particle. No
    numerically stable integration was attained that covers the whole
    range $0\ldots 1$. Therefore, a 7th order series expansion of the
    dissipation function $\phi(\tau)$ in $\tau^{-1}$ was employed to
    generate the correct asymptotic behaviour as $\tau_0^{-2}\to 0$,
    while the full expression for $\phi(\tau)$ was evaluated
    numerically for the other branch. As a guide to the eye, a
    matching curve has been added by hand. }
    \label{fig:comparison_Orrit_relaxation_times}
\end{figure}

Beyond the above limiting results for spherical particles, analytical
estimates for the rotational hot Brownian motion of anisotropic
particles can readily be obtained in the slender rod limit. To find
the friction coefficient per unit length for an infinitely long hot
cylinder, \fref{eq:Stokes_eq} has to be extended by adding Oseen's
term $\rho\vec{U}\cdotp\nabla\vec{u}$ to the force on the right hand
side \cite{Batchelor:2000}. The effective friction coefficient for
stationary translation along the main axis can then be calculated as
\cite{RingsNote}
\begin{equation}
  \label{eq:friction_coeff_cylinder_parallel}
  \bar{\zeta}_\mathrm{HBM}^\parallel=2\pi\left(\int_R^\infty\frac{e^{-\int_R^r\frac{\rho
          U}{\eta(\tilde{r})}\,\rmd\tilde{r}}}{\eta(r) r}\,\rmd r\right)^{-1},
\end{equation}
where the radius $r^\ast$ now denotes the distance to the symmetry
axis of the cylinder. Exploiting the formal analogy between the
Reynolds number $\rho U R/\eta(R)$ and the aspect ratio $L/(2R)$ in
cutting off the hydrodynamic divergences under isothermal conditions,
we arrive at a plausible estimate for the friction coefficient of a
slender cylinder of finite length $L$ and radius $R$, namely,
\begin{equation}
  \label{eq:zeta_rod_approx}
  \bar{\zeta}_\mathrm{HBM}^\parallel=2\pi\eta(R)\bigl(\ln[\eta(R)L/(2\eta_0 R)]-\gamma\bigr)^{-1}.
\end{equation}
For transverse and rotational motion, expressions corresponding to
\fref{eq:friction_coeff_cylinder_parallel} are more complicated to
calculate, due to the lower symmetry of the temperature and fluid
velocity fields. However, simple estimates for the effective
transverse and rotational friction coefficients are readily obtained
if one neglects a potential dependence of the ratios of the various
friction coefficients on the heating, namely,
\begin{equation}
  \label{eq:zeta_perp}
  \bar{\zeta}_\mathrm{HBM}^\perp \approx 2 \bar{\zeta}_\mathrm{HBM}^\parallel
  \text{ and } \bar{\zeta}_\mathrm{HBM}^\mathrm{r}\approx
  (L^2/12)\bar{\zeta}_\mathrm{HBM}^\perp\,.
\end{equation}
The result for $\bar{\zeta}_\mathrm{HBM}^\parallel$, normalized to its
isothermal limit, is depicted in the left panel of
\fref{fig:comparison_Orrit_relaxation_times}, for various aspect
ratios.  For aspect ratios that are large enough to admit the
slender-rod approximation, the temperature dependence of the
expression \eqref{eq:zeta_rod_approx}, normalized to its isothermal
limit, is close to the normalized rotational friction coefficient of a
sphere.

A practical estimate for $\THBM$ of a hot rotating rod is obtained to
first order in $\Delta T$ by ignoring the temperature dependence of
the viscosity and treating the rod as a prolate spheroid. In this
case, the flow field $\vec{u}(\vec{r})$ is known analytically
\cite{Chwang:1975}. It is constructed by a line distribution of some
moments of the fundamental solutions to the Stokes equations under
different singular forcings, so-called stresslets, rotlets, and
potential quadrupoles. From $\vec{u}(\vec{r})$, a straightforward
calculation yields the ``radial'' dissipation function $\phi(\tau)$
from which $\THBM$ is calculated for a given temperature field
$T(\vec{r})$ around the spheroid. In spheroidal coordinates,
$(\tau,\zeta,\varphi)$, where $x=c\tau\zeta$,
$y=c\sqrt{(\tau^2-1)(1-\zeta^2)}\cos\varphi$,
$z=c\sqrt{(\tau^2-1)(1-\zeta^2)}\sin\varphi$, the temperature field
\begin{equation}
  \label{eq:T_spheroid}
  T(\tau)=T_0+\Delta T \mathrm{arccot}\,\tau / \mathrm{arccot}\,\tau_0\,
\end{equation}
obtained from Fourier's law for constant heat conductivity only
depends on the ``radius'' $\tau$ and the eccentricity
$\tau_0^{-2}=1-(2R/L)^2$ of the particle.  Using it together with the
isothermal $\phi(\tau)$ in \fref{eq:T_HBM}, we obtain the first order
term of the series expansion of $\THBM$ for slender particles in
powers of $\Delta T$. Its coefficient $(\THBM-T_0)/\Delta T$ is
plotted in \fref{fig:comparison_Orrit_relaxation_times}. Deviations
from the value $3/4$ for a sphere, \fref{eq:approx_THBM}, only become
apparent for extreme aspect ratios $L/(2R) > 10^2$, corresponding to
eccentricities $\tau_0^{-2}>0.9999$.

Note that the high power of the radial distance $r$ in the denominator
under the integral in \eqs{\ref{eq:zeta_final}) \&
  (\ref{eq:T_HBM_rot_final}} suggests that the effective rotational
temperature and friction should be higher and lower than their
translational counterparts, respectively. This conclusion is supported
by recently published experimental data for gold nanorods, where the
effective temperature for their rotational Brownian motion in an
optical trap is ``found to be close to the particle's temperature''
\cite{Ruijgrok:2011}. The plots of our predictions in
\fref{fig:comparison_Orrit_relaxation_times} suggest that our results
for a spherical particle should still provide a reasonably good
approximation for these rods, which have an aspect ratio of about
two. To make closer contact with the experiments, we have used
\fref{eq:zeta_final} and the differential shell method
\cite{Rings:2011} to calculate the rotational and translational
friction coefficient for the viscosity field of \fref{eq:VF}. The
inset of \fref{fig:comparison_Orrit_relaxation_times} depicts the
ratio of the estimated rotational and translational relaxation times
$\tau_\mathrm{r,t}\propto\zeta_\mathrm{HBM}^\mathrm{r,t}/\Delta T$ for
the experimental particle in the optical trap, assuming a linear
dependence between the trap stiffness and the heating $\Delta T$. The
prediction is seen to be in good qualitative agreement with Fig.~3 of
Ref.~\cite{Ruijgrok:2011}.


One may ask, whether any more general statements can be made
concerning the relative magnitude of the various effective
temperatures of hot Brownian motion introduced so far, independent of
the shape of the particle, and based on some generic material
properties (e.g.\ that the solvent viscosity decreases upon heating).
In particular, one expects that, under otherwise identical conditions,
the effective temperature $\THBM$ will usually be higher for
rotational than for translational motion, because the solvent velocity
field around the particle is more localized near the hot particle for
rotation than for translation.  This suggests that the temperature
ordering
\begin{equation}
  T_0 \leq T_\mathrm{HBM}^\mathrm{t} \leq \THBM  \leq T_\mathrm{s}
  \leq T_\mathrm{p} 
\end{equation}
might be quite generic ($T_\mathrm{p}$ and $T_\mathrm{s}$ are the
particle temperature and the solvent temperature at the particle
surface, respectively, $T_0$ is the ambient temperature). The claim is
corroborated by the results for a temperature-independent viscosity
and for a viscosity step (see appendix), and also by experimental
observations \cite{Rings:2010,Ruijgrok:2011}.
\section{MD simulations}
Numerical simulations allow for a more accurate check of some of our
theoretical prediction than the quoted experiments, since we can
better control the ``experimental'' conditions. Our simulations of
rotational hot Brownian motion are based on the same setup as in the
translational case \cite{Chakraborty:2011}. In brief, the system is
modelled as a Lennard-Jones fluid with a radial pair potential
\mbox{$V(r)=4 \epsilon [(\sigma/r)^{12}-(\sigma/r)^6]$}. While 107233
particles are treated as solvent, a spherical cluster of 767
particles, additionally interconnected by a FENE potential
\mbox{$V(r)=-0.5 \kappa R_0^2 \log[1-(r/R_0)^2]$} with $\kappa=30
\epsilon/\sigma^2$, $R_0=1.5\sigma$, forms the nanoparticle.
\begin{figure}
  \centering
  \includegraphics[width=0.6\linewidth]{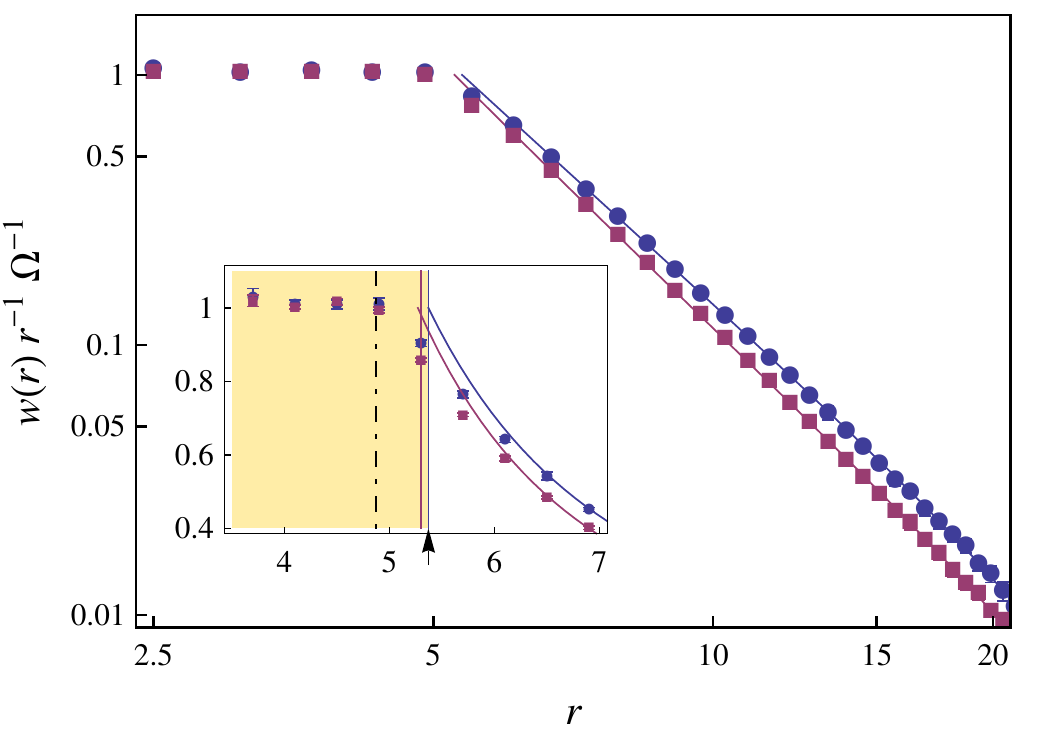}
  \caption{Radial variation of the measured angular velocity $\w(r)/r$
    normalized by the angular velocity of the hot nanoparticle for
    $T_\mathrm{p}=0.9 \epsilon/k_\mathrm{B}$ ({\color{myblue} {\large
        $\bullet$}}) and $1.25 \epsilon/k_\mathrm{B}$
    ({\color{mypurple}\footnotesize $\blacksquare$}). The solid lines
    are the corresponding plots of \fref{eq:gensol_gen_problem}
    evaluated using the measured viscosity and temperature
    profiles. \emph{Inset:} The variation of the angular velocity near
    the nanoparticle surface. The vertical solid lines indicate the
    positions of the hydrodynamic boundary condition for the predicted
    hydrodynamic flow fields \eqref{eq:gensol_gen_problem} fitted to
    the simulation data. The shaded region (its extension marked by an
    arrow) indicates the equivalent sphere radius for the nanoparticle
    (dot-dashed line) plus $\sigma/2$. The comparison reveals a weak
    apparent slip at high nanoparticle temperatures due to the radial
    solvent density variation induced by the heating.}
  \label{fig:flowfield}
\end{figure}
Each simulation run consists of an initial isothermal equilibration
phase at \mbox{$T=0.75\,\epsilon/k_\mathrm{B}$}, followed by a heating
phase, where the nanoparticle is kept at a constant elevated
temperature $T_\mathrm{p}$, and a boundary layer at the periphery of
the simulation volume is kept at $T=0.75\,\epsilon/k_\mathrm{B}$. All
measurements were performed once a stationary state had been attained
in the heating phase. The results were averaged over 30 independent
trajectories of $2\times 10^5$ steps (corresponding to a physical
duration of $1\,$ns). For further details concerning the simulation
method, we refer the reader to Ref.~\cite{Chakraborty:2011}

The rotational diffusion coefficient $\DHBM$ was determined from the
decay of the autocorrelation function of an orientation vector
assigned to the freely diffusing hot nanoparticle
\fref{fig:artistic_sketch}. The rotational friction coefficient
$\zetaHBM$ was determined by applying a constant external torque to
the nanoparticle and measuring its angular velocity. To gain more
insight into the microscopic details, we also recorded the radially
varying angular velocity in concentric shells of thickness $\sigma/5$
around the nanoparticle, in this case.  \Fref{fig:flowfield} shows the
excellent agreement of the measured angular velocity field $\w(r)/r$
with the theoretical prediction in \fref{eq:gensol_gen_problem} --
except in the close vicinity of the nanoparticle (inset of
\fref{fig:flowfield}) where solvent density variations are prominent,
which have been neglected in the analytical calculation.  For a
comparison with the predictions, we therefore treated the radius $R$
of the nanoparticle, which affects $\w(r)$ via the boundary condition,
as a free parameter. When the heating of the nanoparticle is small,
$R$ agrees with the value $R_\mathrm{H} +\sigma/2$, with the radius
$R_\mathrm{H}$ of an equivalent solid sphere given by
$R^2_\mathrm{H}=\frac{5}{6N^2}\sum_{i,j=1}^{N}(\vec{r}_i-\vec{r}_j)^2$
\cite{Vladkov:2006,Heyes:1998}. For strong heating, $R$ decreases
slightly, due to the mentioned solvent density variation, which
results in a weak surface slip of the hydrodynamic flow field (inset
of \fref{fig:flowfield}).

Finally, we pursued three routes for determining the hot
nanoparticle's effective temperature: \emph{(i)} by measuring the
rotational diffusion and friction coefficients, $\DHBM$ and
$\zetaHBM$, and making use of the generalized Einstein relation in
\fref{eq:GER} ({\large \color{myblue} $\bullet$} in \fref{fig:fig4});
\emph{(ii)} by fitting the recorded angular distribution $p(\theta)$
in a harmonic angular confinement potential
$V(\theta)=\frac{K}{2}(\theta-\pi/2)^2$ with a Boltzmann distribution
$p(\theta ) \propto \exp(-V(\theta)/k_B\THBM)$
\mbox{({\color{myblue}\footnotesize $\square$} in \fref{fig:fig4})};
and \emph{(iii)} by measuring the viscosity and temperature profiles,
$\eta(r)$ and $T(r)$ and calculating $\THBM$ via
\fref{eq:T_HBM_rot_final} (solid line in \fref{fig:fig4}). All three
methods agree within the error bars, thereby confirming our
theoretical predictions for the effective friction $\zetaHBM$, the
effective temperature $\THBM$, and the generalized Einstein relation
of rotational hot Brownian motion, \fref{eq:GER}.


\section{Conclusions}
We have analyzed the rotational Brownian motion of a heated
(nano-)particle in an otherwise unheated solvent. Particles of
spherical, spheroidal and slender cylindrical shape have been
considered.  While we have obtained exact results for the effective
rotational friction and temperature in the case of spherical
particles, we contended ourselves with approximate estimates for
prolate shapes. Disregarding a potential temperature dependence of the
ratio between the transverse and the longitudinal friction for a rod
in \fref{eq:zeta_perp} is justified for large particle aspect ratios,
since the near-field solutions for the solvent flow are dominated by
terms with the same radial dependence. Also, while Oseen's treatment
of the infinite cylinder in \fref{eq:friction_coeff_cylinder_parallel}
adds a nonlinearity in $\vec{U}$ to the Stokes problem, calling for a
nonlinear (and non-Markovian) theory, the linear analysis of
Ref.~\cite{Chakraborty:2011} remains valid for finite particles.  It
leaves only the (possibly difficult) technical problem of evaluating
the flow velocity field from which the effective parameters are
calculated.

We validated our theoretical predictions against large-scale molecular
dynamics simulations of a Lennard-Jones system with an immersed hot
colloid, and by comparison with recent measurements of diffusing hot
nanorods \cite{Ruijgrok:2011}. Thereby, we could quantitatively
establish an effective equilibrium description for the nonequilibrium
rotational motion of a heated Brownian particle. While the general
procedures and results resemble those for translational motion
\cite{Rings:2011}, their analytical tractability and their detailed
structure was found to be substantially different. From a comparison
of the rotational and translational case, an ordering of the effective
temperatures governing hot Brownian motion emerged. Our findings
provide the basis for a quantitative investigation of the Brownian
dynamics of particles of anisotropic shapes, interactions between hot
particles \cite{Golestanian:2012}, and self-thermophoretic Janus
particles that move on persistent Brownian paths determined by a
mutual competition of rotational and translational hot Brownian motion
\cite{Jiang:2010}.

\ack This work was supported by the Deutsche Forschungsgemeinschaft
(DFG) via FOR 877, and by the Alexander von Humboldt foundation.

\section*{References}
\bibliography{library}
\bibliographystyle{unsrt}

\appendix
\section*{Appendix: Stokes' problem for a viscosity step.}
\setcounter{section}{1} From equations (49) and (53) of
Ref.~\cite{Rings:2011}, the viscous dissipation function
$\phi(r)=\int\phi(\vec{r})\,\rmd\theta\rmd\varphi$ for a translating
sphere, driven by the constant force $\vec{F}=F\vec{e}_z$, follows as
\begin{eqnarray}
  \label{eq:visc_dissip_density}
  \phi(r) &= 
  \frac{4\pi\eta}{r^8}\left[15 a_3^2-3F a_3 r^2/(2\pi\eta)\right.\nonumber\\
   &\left.\phantom{= }+r^4 \left(F^2/(16\pi^2\eta^2)+ F/(\pi\eta) a_2
       r^3+10 a_2^2 r^6\right)\right],
\end{eqnarray}
which depends on $\eta(r)$ in an intricate way, namely via equations
(44) of Ref.~\cite{Rings:2011}. While this allows for an explicit
computation of $T_\mathrm{HBM}^\mathrm{t}$ once $\eta(r)$ is
specified, general statements about the relative magnitudes of
$T_\mathrm{HBM}^\mathrm{t}$ and $\THBM$ seem thus hard to attain, even
for the highly symmetric case of a spherical bead. We therefore
contend ourselves with outlining the principles how such general
considerations might proceed, namely by considering a spherical bead
and the class of step profiles,
\begin{equation}
  \label{eq:step_visc}
  r\le\beta R:\;\eta(r)=\eta_0/\kappa\;,\quad r>\beta
  R:\;\eta(r)=\eta_0\;, 
\end{equation}
for which both the translational and rotational effective temperatures
can be calculated analytically. Note that the common case of a fluid
viscosity that decreases upon heating corresponds to $\kappa>1$, while
$\kappa<1$ holds for dilute gases.

The effective rotational temperature $\THBM$ follows from
\fref{eq:T_HBM_rot_final} and takes the simple form
\begin{equation}
  \label{eq:T_HBM_rot_step}
  \frac{\THBM-T_0}{\Delta T R} = \frac{3 \left(\beta ^4-1\right) \kappa +3}{4 \left(\beta ^4
      \kappa -\beta \kappa +\beta \right)}\,,
\end{equation}
where $T_0$ is the ambient temperature.

For the translational case, the general ansatz for the velocity and
pressure fields, $\vec u=u_r\vec{e}_r+u_\theta\vec{e}_\theta$ and $p$,
reads \cite{LifPit:2004}:
\begin{eqnarray}
  \label{eq:ansatz}
    u_r(r,\theta)&=\left(a_0+\frac{a_1}{r}+a_2r^2+\frac{a_3}{r^3}\right)\cos\theta\\
    u_\theta(r,\theta)&=-\left(a_0+\frac{a_1}{2r}+2a_2r^2-\frac{a_3}{2r^3}\right)\sin\theta\\
    p(r,\theta)&=p_0+\left(\frac{a_1}{r^2}+10a_2r\right)\eta\cos\theta\;,
\end{eqnarray}
where $u_r$ and $u_\theta$ denote the radial and polar velocity
components, respectively. While the coefficients are
\cite{LifPit:2004}
\begin{equation}
  \label{eq:const_coeffs}
  a_0=u_0,\quad a_1=-3u_0R/2,\quad a_2=0,\quad a_3=u_0R^3/2
\end{equation}
in an infinite homogeneous system, where $u_0$ denotes the particle
velocity relative to the resting fluid at infinity, the coefficients
$a_i$ each take two $r$-dependent values for the step profile of the
viscosity. These coefficients are found by matching the boundary
conditions for the velocity and the stress at $r=R$, $r=\beta R$, and
$r\to\infty$. More precisely, the following linear system of equations
needs to be solved:
\begin{equation}
  \label{eq:lin_system}
  \mathrm{M}\vec{a}=\vec{v}
\end{equation}
with
\begin{eqnarray}
  \label{eq:lin_system_abbrevs}
  \fl\mathrm{M}\equiv\left(
    \begin{array}{ccccccc}
      0 & 0 & 0 & 0 & 1 & 0 & 0 \\
      R^3 & R^2 & R^5 & 1 & 0 & 0 & 0 \\
      -2 R^3 & -R^2 & -4 R^5 & 1 & 0 & 0 & 0 \\
      R^3 \beta ^3 & R^2 \beta ^2 & R^5 \beta ^5 & 1 & -R^3 \beta ^3
      & -R^2 \beta ^2 & -1 \\
      -2 R^3 \beta ^3 & -R^2 \beta ^2 & -4 R^5 \beta ^5 & 1 & 2 R^3
      \beta ^3 & R^2 \beta ^2 & -1 \\
      0 & \frac{R^2 \beta ^2 \eta_0}{\kappa } & \frac{2 R^5
        \beta ^5 \eta_0}{\kappa } & \frac{2 \eta_0}{\kappa } & 0 & -R^2 \beta ^2 \eta_0 & -2
      \eta_0 \\
      0 & 0 & \frac{R^5 \beta ^5 \eta_0}{\kappa } &
      \frac{\eta_0}{\kappa } & 0 & 0 & -\eta_0
    \end{array}
  \right),\\
  \hfill\vec{a}\equiv\left(
    \begin{array}{c}
      a_0(r<\beta R)\\
      a_1(r<\beta R)\\
      a_2(r<\beta R)\\
      a_3(r<\beta R)\\
      a_0(r>\beta R)\\
      a_1(r>\beta R)\\
      a_3(r>\beta R)
    \end{array}
  \right),\quad\vec{v}\equiv\left(
    \begin{array}{c}
      U\\
      0\\
      0\\
      0\\
      0\\
      0\\
      0
    \end{array}
  \right),\quad a_2(r>\beta R)=0\,.\hfill
\end{eqnarray}
We find as solution,
\begin{equation}
  \label{eq:lin_system_solution}
  \vec{a}=\frac{U}{\mathcal{C}}\left(
    \begin{array}{c}
      \beta\kappa[\beta ^5 (4 \kappa +6)+5 \beta ^2 (\kappa -1)-9 \kappa +9]\\[3pt]
      -3R\beta\kappa[\beta ^5 (2 \kappa +3)-2 \kappa +2]\\[3pt]
      -3\beta\kappa R^{-2}(\beta ^2-1) (\kappa -1)\\[3pt]
      R^3 \beta ^3 \kappa[\beta ^3 (2 \kappa +3)-2 \kappa
      +2]\\[3pt]
      \mathcal{C}\\[3pt]
      -3R\beta[\beta ^5 (2 \kappa +3)-2 \kappa +2]\\[3pt]
      R^3 \beta ^3 [-3 \beta ^5 (\kappa -1)+5 \beta ^3
          \kappa -2 \kappa +2]
    \end{array}
  \right),
\end{equation}
where $\mathcal{C}=4+6 \beta ^5+\left(6 \beta ^6+3 \beta ^5-10 \beta
  ^3+9 \beta -8\right) \kappa+(\beta -1)^4 \left[\beta (4 \beta
  +7)+4\right] \kappa ^2$.\\
We use this solution together with
\fref{eq:visc_dissip_density} to compute
\begin{equation}
  \label{eq:T_HBM_trans_step}
   \frac{T_\mathrm{HBM}^\mathrm{t}-T_0}{\Delta T
     R}=\frac{\int_R^{\beta R}\phi r\,\rmd r+\int_{\beta R}^\infty\phi
     r\,\rmd r}{\int_R^{\beta R}\phi r^2\,\rmd r+\int_{\beta
       R}^\infty\phi r^2\,\rmd r}\;.
\end{equation}
Dividing the expression in \fref{eq:T_HBM_rot_step} by this result and
evaluating the integrals, we obtain
\begin{equation}
  \label{eq:ratioT_rot_T_trans}
  \frac{T_\mathrm{HBM}^\mathrm{r}-T_0}{T_\mathrm{HBM}^\mathrm{t}-T_0}=\frac{p_0+p_1\kappa+p_2\kappa^2+p_3\kappa^3+p_4\kappa^4}{q_0+q_1\kappa+q_2\kappa^2+q_3\kappa^3+q_4\kappa^4}\;,
\end{equation}
where ($\beta\equiv1+\epsilon$)
\begin{eqnarray}
  \label{eq:termwise_comparison}
    p_0-q_0&=8 \left(3 (\epsilon +1)^5+2\right)^2\\
    p_1-q_1&=\epsilon  \bigl(162 \epsilon ^{13}+2178 \epsilon ^{12}+13482 \epsilon ^{11}+51030
      \epsilon ^{10}+131931 \epsilon ^9\\
      &\;+246390 \epsilon ^8+343605 \epsilon ^7+366360 \epsilon
      ^6+304685 \epsilon ^5\\
      &\;+200610 \epsilon ^4+104325 \epsilon ^3+41350 \epsilon ^2+11400 \epsilon +1800\bigr)\\
    p_2-q_2&=3 \epsilon ^2 \bigl(24 \epsilon ^{13}+423
      \epsilon ^{12}+3252 \epsilon ^{11}+14743 \epsilon ^{10}+44630
      \epsilon ^9\\
      &\;+96664 \epsilon ^8+156760 \epsilon ^7+196480
      \epsilon ^6+193820 \epsilon ^5\\
      &\;+150415 \epsilon ^4+89750 \epsilon^3+39325 \epsilon ^2+11550 \epsilon +1800\bigr)\\
    p_3-q_3&=\epsilon
    ^3 \bigl(96 \epsilon ^{12}+1251 \epsilon ^{11}+7704 \epsilon
    ^{10}+30021 \epsilon ^9+83280 \epsilon ^8\\
    &\;+174165 \epsilon^7+281090 \epsilon ^6+349995 \epsilon ^5+330540
    \epsilon ^4\\
    &\;+229950\epsilon ^3+112500 \epsilon ^2+35100 \epsilon+5400\bigr)\\
    p_4-q_4&=2 \epsilon ^8 \bigl(16 \epsilon ^7+159 \epsilon ^6+681 \epsilon ^5+1634 \epsilon ^4+2400 \epsilon ^3+2220
    \epsilon ^2\\
    &\;+1260 \epsilon +360\bigr)\;.
\end{eqnarray}
Obviously, since $\beta>1$ by definition, and thus $\epsilon>0$, each
coefficient $p_i$ in the numerator of \fref{eq:ratioT_rot_T_trans} is
strictly larger than the corresponding coefficient in the denominator,
$q_i$. Hence, for a viscosity profile that increases from $\eta(R)$ to
$\eta_0$ in a single step, we have shown that
$T_\mathrm{HBM}^\mathrm{r}>T_\mathrm{HBM}^\mathrm{t}$ holds.

\end{document}